\documentclass{article}

\usepackage{color}
\usepackage{amsmath, amssymb}

\newcommand{\ZM}{\mathbb{Z}}

\newcommand{\PM}{\mathbb{P}}

\newcommand{\bec}[1]{\mbox{\boldmath $#1$}}
\newcommand{\miniket}[1]{\vert#1\rangle}

\begin{document}

\title{{\bf A quantization of interacting particle systems}
\vspace{30mm}}

\author{Jir\^o AKAHORI, $\quad$ Norio KONNO$^{\ast}$, $\quad$ Rikuki OKAMOTO$^{\ast \ast}$ \\ \\
Department of Mathematical Sciences \\
College of Science and Engineering \\
Ritsumeikan University \\
1-1-1 Noji-higashi, Kusatsu, 525-8577, JAPAN \\
e-mail: akahori@se.ritsumei.ac.jp, \  n-konno@fc.ritsumei.ac.jp$^{\ast}$, \\ 
ra0099vs@ed.ritsumei.ac.jp$^{\ast \ast}$ \\
\\
\\
Iwao SATO \\ \\
Oyama National College of Technology \\
771 Nakakuki, Oyama 323-0806, JAPAN \\
e-mail: isato@oyama-ct.ac.jp 
}

\date{\empty }

\maketitle

\vspace{50mm}


\vspace{20mm}


\begin{small}
\par\noindent
\\
\end{small}









\clearpage

\begin{abstract}
Interacting particle systems studied in this paper are probabilistic cellular automata with nearest-neighbor interaction including the Domany-Kinzel model. A special case of the Domany-Kinzel model is directed percolation. We regard the interacting particle system as a Markov chain on a graph. Then we present a new quantization of the interacting particle system. After that, we introduce a zeta function of the quantized model and give its determinant expression. Moreover, we calculate the absolute zeta function of the quantized model for the Domany-Kinzel model. 
\end{abstract}

\vspace{10mm}

\begin{small}
\end{small}

\vspace{10mm}

\section{Introduction \label{sec01}}
The interacting particle system (IPS) considered here is the probabilistic cellular automaton (PCA) with nearest-neighbor interaction on $\mathbb{P}_N = \{ 0,1, \ldots, N-1 \}$. There are two states ``0" or ``1" at each location. Let $\eta (x) \in \{0,1\}$ be the state of the location $x \in \PM_N$, i.e., $x=0,1,\ldots, N-1$. The set of configurations is $\{0,1\}^{\mathbb{P}_N}$ with $2^N$ elements. For example, if $N=3$, then a configuration $(1,0,1) \in \{0,1\}^{\mathbb{P}_3}$ means that the state``1" at location 0, the state ``0" at location 1, and the state ``1" at location 2. In other words, $(\eta (0), \eta (1), \eta (2)) = (1,0,1)$. Remark that we sometimes abbreviate $101$ instead of $(1,0,1)$. The IPS includes the Domany-Kinzel model (DK model) determined by two parameters $p$ and $q$ with $p, \ q \in [0,1]$, which was introduced in \cite{DomanyKinzel1984}. A special case of the DK model is directed percolation (DP). It is also called oriented percolation. Then bond (resp. site) DP is defined by $q=2p - p^2$ (resp. $q=p$). As for DP, see \cite{Durett1988, Lubeck2006, HHLP2008}, for example.

Let $G = (V(G), E(G))$ be a graph, where $V(G)$ the set of vertices and $E(G)$ is the set of nonoriented edges $uv$ joining two vertices $u$ and $v$. One of the typical quantum walks on $G$ is considered as a quantization of the random walk jumping to one of the nearest-neighbor vertices, see \cite{Konno2008, Venegas, Portugal}, for example. In the present paper, we propose a new quantization of our IPS on a graph $G$ by regarding a configuration in $\{0,1\}^{\mathbb{P}_N}$ as a vertex in $G$. Then the IPS can be considered as a Markov chain on a graph $G$, where $G = G^{(1)} \cup G^{(2)}$, i.e., $G$ is a disjoint union of $G^{(1)}$ and $G^{(2)}$. Concerning $V(G^{(i)}) \ (i=1,2)$ for $N \ge 2$, we put
\begin{align*} 
V(G^{(1)}) 
&= \{ \ast_0 \ast_1 \ldots \ast_{N-2} 0 \ \mid \ \ast_k \in \{0,1\} \ (k=0,1, \ldots, N-2) \},
\\
V(G^{(2)}) 
&= \{ \ast_0 \ast_1 \ldots \ast_{N-2} 1 \ \mid \ \ast_k \in \{0,1\} \ (k=0,1, \ldots, N-2) \}.
\end{align*}
For example, if $N=2$, then
\begin{align*} 
V(G^{(1)}) = \{ 00, 10 \}, \qquad V(G^{(2)}) = \{ 01, 11 \},
\end{align*}
and if $N=3$, then
\begin{align*} 
V(G^{(1)}) = \{ 000, 010, 100, 110 \}, \qquad V(G^{(2)}) = \{ 001, 011, 101, 111 \}.
\end{align*}
Thus we see $|V(G^{(1)})| = |V(G^{(2)})| = 2^{N-1}$, where $|A|$ is the number of elements in a set $A$. Moreover, each $G^{(i)} \ (i=1,2)$ is $K_{2^{N-1}}$ with $2^{N-1}$ loops, where $K_n$ is the complete graph with $n$ vertices and $n(n-1)/2$ nonloop edges. Thus, each $G^{(i)}$ has $2^{N-1}$ vertices, $2^{N-2} (2^{N-1}-1)$ nonloop edges and $2^{N-1}$ loops. We should note that $G^{(i)}$ has one loop at each vertex of $G^{(i)}$ for $i=1,2$. Therefore, we confirm that $G = G^{(1)} \cup G^{(2)}$ has $2^{N}$ vertices, $2^{N-1} (2^{N-1}-1)$ nonloop edges and $2^{N}$ loops. See Section~\ref{sec03} for details. Let ${\bf P}^{(1)}$ and ${\bf P}^{(2)}$ be transposed transition matrices of Markov chains on $G^{(1)}$ and $G^{(2)}$, respectively. Remark that ${\bf P}^{(1)}$ and ${\bf P}^{(2)}$ are $2^{N-1} \times 2^{N-1}$ matrices. Furthermore, $2^{N} \times 2^{N}$ transposed transition matrix ${\bf P}$ for the IPS on $G = G^{(1)} \cup G^{(2)}$ is expressed as ${\bf P} = {\bf P}^{(1)} \oplus {\bf P}^{(2)}$. 

The quantum walk, i.e., quantization of the IPS, derived from our quantization is defined by two different quantum walks on the above-mentioned graph $G$. Then, the time-evolution matrices ${\bf U}^{(1)}$ and ${\bf U}^{(2)}$ for the quantum walks correspond to ${\bf P}^{(1)}$ and ${\bf P}^{(2)}$, respectively. Here ${\bf U}^{(1)}$ and ${\bf U}^{(2)}$ are $2^{2(N-1)} \times 2^{2(N-1)}$ unitary matrices. Furthermore, $2^{2N-1} \times 2^{2N-1}$ time-evolution matrix ${\bf U}$ on $G = G^{(1)} \cup G^{(2)}$ is given by ${\bf U} = {\bf U}^{(1)} \oplus {\bf U}^{(2)}$ corresponding to ${\bf P} = {\bf P}^{(1)} \oplus {\bf P}^{(2)}$.

Next we consider a finite connected graph $G$ with $n$ vertices, $m$ nonloop edges and $n$ loops, where $G$ has one loop at each vertex of $G$. Note that we consider this $G$ as above-mentioned $G^{(i)}$ for each $i=1,2$. Thus, $G$ is connected. Moreover, $D_G$ denotes the symmetric digraph of $G$, where $D_G$ has one diloop at each vertex.  Let $D(G)$ be the arc set of $D_G$. Furthermore, let $p : D(G) \to [0,1]$ be a transition probability on $D(G)$. Then $n \times n$ transposed transition matrix ${\bf P}$ is determined by $p$. Thus we construct a Markov chain $(D_G, p)$ on $G$. For a Markov chain $(D_G, p)$ on $G$, we introduce a new quantization of $(D_G, p)$. In fact, each component of $(n+2m) \times (n+2m)$ unitary matrix ${\bf U}$ is given by $p$. Then we call ${\bf U}$  a quantization of $(D_G, p)$. 


This ${\bf U}$ can be expressed as ${\bf U} = 2 {\bf K} {\bf L}^{T} - {\bf J}$, where ${\bf K}$ and ${\bf L}$ are $(n+2m) \times n$ matrices and ${\bf J}$ is an $(n+2m) \times (n+2m)$ matrix. Here $A^{T}$ is the transepose of a matrix $A$. From this expression, we show that ${\bf U}$ is unitary (Proposition 1). Moreover, we present a zeta function $\zeta (G, {\bf U}, u)$ of $G$ for ${\bf U}$ as follows:
\begin{align*}
\zeta (G, {\bf U}, u) = \det ( {\bf I}_{n+2m} -u {\bf U} )^{-1}. 
\end{align*}
Then, using an $n \times n$ symmetrized matrix ${\bf S}$ for ${\bf P}$, we obtain the following result (Theorem 2): 
\begin{align*}
\zeta (G, {\bf U} ,u)^{-1} = ( 1 + u )^n (1 - u^2)^{m-n} \det \left( ( 1+u^2) {\bf I}_{n} -2 u {\bf S} \right).
\end{align*}
Furthermore, it follows from this that we have the characteristic polynomial which is similar to the Konno-Sato theorem \cite{KonnoSato}:
\begin{align*}
\det ( \lambda {\bf I}_{n+2m} - {\bf U} )=( \lambda +1)^n ( \lambda {}^2 -1)^{m-n} \det  \left( ( \lambda {}^2 +1) {\bf I}_{n} -2 \lambda {\bf S} \right).
\end{align*}
\par
Finally, we treat $N=2$ case for the DK model with parameters $p, \ q \in [0,1]$. The important point is that the zeta function $\zeta (G, {\bf U}, u)$ is an absolute automorphic form. Note that if we consider the corresponding zeta function $\zeta (G, {\bf P}, u) = \det ( {\bf I}_{n} -u {\bf P} )^{-1}$ for ${\bf P}$ instead of ${\bf U}$, then the zeta function $\zeta (G, {\bf P}, u)$ is not necessarily absolute automorphic form. In particular, for each of two cases, $(p,q)=(1/2,0)$ and $(p,q)=(0,1/2)$, we compute absolute zeta functions of our zeta function $\zeta (G, {\bf U} ,u)$. Concerning the absolute zeta function for zeta functions based on the quantum walk and on quantum cellular automata (QCA), see Konno \cite{Konno2023} and Akahori et al. \cite{AkahoriEtAL2023}, respectively. One of the motivations for computing the absolute zeta functions is as follows. The DK model is defined by two parameters $p$ and $q$ with $p, \ q \in [0,1]$. In the $(p,q)$-phase diagram of the DK model for $N = \infty$ case, along the critical line, the model exhibits phase transition, i.e., from active phase to inactive one. The phase transition of the DK model has not yet been sufficiently investigated. In fact, from the mathematical point of view, the critical line of the phase diagram for the DK model is not rigorously known. Therefore, we would like to know some information on the critical line by using the functional equation of the absolute zeta function for the zeta function of our quantization of the DK model.

The rest of this paper is organized as follows. Section~\ref{sec02} presents the definition of the IPS on $\mathbb{P}_N$. In addition, we introduce the DK model as a special case. Section~\ref{sec03} deals with a new quantization of our IPSs. In Section~\ref{sec04}, we propose a zeta function $\zeta (G, {\bf U}, u)$. After that, we give a determinant expression of $\zeta (G, {\bf U}, u)$ and obtain eigenvalues of ${\bf U}$. Section~\ref{sec05} treats a quantized model for the DK model. In Section~\ref{sec06}, we review the absolute zeta function and its related results. Moreover, we calculate the absolute zeta function of a quantized model of the DK model for $N=2$ case. Section~\ref{sec07} is devoted to summary.

\section{Definition of IPS \label{sec02}}
According to the notation of our previous papers \cite{AkahoriEtAL2023, KomatsuEtAl2022, KiumiEtAl2022}, this section gives the definition of our IPSs on $\mathbb{P}_N = \{ 0,1, \ldots, N-1 \}$ with $N \ge 2$: $\mathbb{Z}$ is the set of integers, $\mathbb{Z}_{\ge}$ is the set of non-negative integers, $\mathbb{Z}_{>}$ is the set of positive integers, $\mathbb{R}$ is the set of real numbers, and $\mathbb{C}$ is the set of complex numbers. 

There are two states ``0" or ``1" at each location for the IPS. Let $\eta (x) \in \{0,1\}$ denote the state of the location $x \in \PM_N$, i.e., $x=0,1,\ldots, N-1$. The set of configurations is $\{0,1\}^{\mathbb{P}_N}$ with $2^N$ elements. In this paper, we put
\begin{align}
\miniket 0 =
\begin{bmatrix}
1 \\
0
\end{bmatrix} 
,
\quad 
\miniket 1 =
\begin{bmatrix}
0 \\
1
\end{bmatrix}
.
\label{zeroone} 
\end{align}
For example, when $N=3$, a configuration $(0,1,1) \in \{0,1\}^{\mathbb{P}_3}$ means the state``0" at location 0, the state ``1" at location 1, and the state ``1" at location 2. That is, $(\eta (0), \eta (1), \eta (2)) = (0,1,1)$. Note that we sometimes abbreviate $011$ instead of $(0,1,1)$. We also write $(0,1,1)$ by $\miniket 0 \miniket 1 \miniket 1 = \miniket 0 \otimes \miniket 1 \otimes \miniket 1 = \vert 011 \rangle$. By using Eq. \eqref{zeroone}, we have
\begin{align*}
\miniket 0 \miniket 1 \miniket 1 = 
\begin{bmatrix}
1 \\
0
\end{bmatrix} 
\otimes
\begin{bmatrix}
0 \\
1
\end{bmatrix} 
\otimes
\begin{bmatrix}
0 \\
1
\end{bmatrix} 
=
\begin{bmatrix}
0 \\
0 \\
0 \\
1 \\
0 \\
0 \\
0 \\
0 
\end{bmatrix}
.
\end{align*}

To define our model, we introduce the {\em local} operator $Q^{(\ell)}$ and the {\em global} operator $Q^{(g)}_N$ in the following way. This definition is based on Katori et al. \cite{KatoriEtAl2004}.

We first define the $4 \times 4$ matrix $Q^{(\ell)}$ by
\begin{align*}
Q^{(\ell)}
=
\begin{bmatrix}
a^{00}_{00} & a^{01}_{00} & a^{10}_{00} & a^{11}_{00} \\ 
a^{00}_{01} & a^{01}_{01} & a^{10}_{01} & a^{11}_{01} \\ 
a^{00}_{10} & a^{01}_{10} & a^{10}_{10} & a^{11}_{10} \\ 
a^{00}_{11} & a^{01}_{11} & a^{10}_{11} & a^{11}_{11}  
\end{bmatrix} 
,
\end{align*}
where $a^{ij}_{kl} \in \mathbb{C}$ for $i,j,k,l \in \{0,1\}.$  Let $\eta_n (x) \in \{0,1\}$ denote the state of the location $x \in \PM_N$ at time $n \in \ZM_{\ge}$. The element of $Q^{(\ell)}$, $a^{ij}_{kl}$, means a transition weight from $(\eta_n (x), \eta_n (x+1)) =(i,j)$ to $(\eta_{n+1} (x), \eta_{n+1} (x+1))=(k,l)$ for any $x=0,1, \ldots, N-2$ and $n \in \ZM_{\ge}$. We call ``$x$" the {\em left site} and ``$x+1$" the {\em right site}. Throughout this paper, we assume that $a^{ij}_{kl}=0$ if $j \not=l$. In words, after the time transition, the state of the right site does not change. Therefore, under this assumption, $Q^{(\ell)}$ is rewritten as
\begin{align*}
Q^{(\ell)}
=
\begin{bmatrix}
a^{00}_{00} & \cdot & a^{10}_{00} & \cdot \\ 
\cdot & a^{01}_{01} & \cdot & a^{11}_{01} \\ 
a^{00}_{10} & \cdot & a^{10}_{10} & \cdot \\ 
\cdot & a^{01}_{11} & \cdot & a^{11}_{11}  
\end{bmatrix} 
,
\end{align*}
where $\cdot$ means 0. In particular, if $a^{ij}_{kl} \in \{0,1\}$, then the IPS is called cellular automaton. Note that if we take $N \to \infty,$ then we should consider the finite size effect. Next we define the $2^N \times 2^N$ matrix $Q^{(g)}_N$ by
\begin{align*}
Q^{(g)}_N
&= \left( {\bf I}_2 \otimes {\bf I}_2 \otimes \cdots \otimes {\bf I}_2 \otimes Q^{(\ell)} \right) \left( {\bf I}_2 \otimes {\bf I}_2 \otimes \cdots \otimes Q^{(\ell)} \otimes {\bf I}_2 \right) 
\\
& \qquad \cdots \left( {\bf I}_2 \otimes Q^{(\ell)} \otimes \cdots \otimes {\bf I}_2 \otimes {\bf I}_2 \right) \left( Q^{(\ell)} \otimes {\bf I}_2 \otimes \cdots \otimes {\bf I}_2 \otimes {\bf I}_2 \right),
\end{align*}
where $\otimes$ is the tensor product and ${\bf I}_n$ is the $n \times n$ identity. For example, if $N=3$, then the $2^3 \times 2^3$ matrix $Q^{(g)}_3$ is 
\begin{align*}
Q^{(g)}_3 
= \left( {\bf I}_2 \otimes Q^{(\ell)} \right) \left( Q^{(\ell)} \otimes {\bf I}_2 \right).
\end{align*}
If $N=4$, then the $2^4 \times 2^4$ matrix $Q^{(g)}_4$ is 
\begin{align*}
Q^{(g)}_4 
= \left( {\bf I}_2 \otimes {\bf I}_2 \otimes Q^{(\ell)} \right) \left( {\bf I}_2 \otimes Q^{(\ell)} \otimes {\bf I}_2 \right) \left( Q^{(\ell)} \otimes {\bf I}_2 \otimes {\bf I}_2 \right).
\end{align*}
For example, when $N=4$, a transition weight from $(\eta_n (0), \eta_n (1), \eta_n (2), \eta_n (3)) =(i_0,i_1,i_2,i_3) \in \{0,1\}^4$ to $(\eta_{n+1} (0), \eta_{n+1} (1), \eta_{n+1} (2), \eta_{n+1} (3)) =(k_0,k_1,k_2,k_3) \in \{0,1\}^4$ is $a^{i_0 i_1}_{k_0 k_1} a^{i_1 i_2}_{k_1 k_2} a^{i_2 i_3}_{k_2 k_3}$ for any $n \in \ZM_{\ge}$.

The above-mentioned model is called the IPS in this paper. We consider two typical classes, one is PCA and the other is QCA. Note that PCA is also called stochastic CA. 

A model in PCA satisfies 
\begin{align*}
a^{00}_{00} + a^{00}_{10} = a^{01}_{01} + a^{01}_{11} = a^{10}_{00} + a^{10}_{10} = a^{11}_{01} + a^{11}_{11} = 1, \quad a^{ij}_{kj} \in [0,1].
\end{align*}
That is, $Q^{(\ell)}$ is a transposed {\em transition matrix} (also called {\em  stochastic matrix}). Furthermore, we easily see that `` $Q^{(\ell)}$ is a transposed transition matrix if and only if $Q^{(g)}_N$ is a transposed transition matrix".

On the other hand, a model in QCA satisfies that $Q^{(\ell)}$ is unitary, i.e., 
\begin{align*}
&
|a^{00}_{00}|^2 + |a^{00}_{10}|^2 = |a^{01}_{01}|^2 + |a^{01}_{11}|^2 = |a^{10}_{00}|^2 + |a^{10}_{10}|^2 = |a^{11}_{01}|^2 + |a^{11}_{11}|^2 = 1,
\\
& 
\qquad a^{00}_{00} \  \overline{a^{10}_{00}} + a^{00}_{10} \  \overline{a^{10}_{10}} =
 a^{01}_{01} \  \overline{a^{11}_{01}} + a^{01}_{11} \  \overline{a^{11}_{11}} =0. 
\end{align*}
This QCA was introduced by Konno \cite{Konno2008b}. As in the case of the PCA, we easily see that `` $Q^{(\ell)}$ is unitary if and only if $Q^{(g)}_N$ is unitary". The QCA was investigated in \cite{AkahoriEtAL2023, KomatsuEtAl2022}. In this paper, we focus on the PCA. We should remark that the above QCA is different from our quantization of PCA defined in Section~\ref{sec03}.

Here we give the DK model as a typical model of PCA. As for the DK model and its related topics, see \cite{DomanyKinzel1984, Durett1988, Lubeck2006, HHLP2008}, for example. The DK model exhibits a phase transition controlled by two parameters $p$ and $q$ with $p, \ q \in [0,1]$. The {\em local} operator $Q^{(\ell)}_{DK}$ is determined by
\begin{align*}
Q^{(\ell)}_{DK}
=
\begin{bmatrix}
1 & \cdot & 1-p & \cdot \\ 
\cdot & 1-p & \cdot & 1-q \\ 
\cdot & \cdot & p & \cdot \\ 
\cdot & p & \cdot & q  
\end{bmatrix} 
\end{align*}
for $p, \ q \in [0,1]$. In particular, if $q=p$ (resp. $q=1-(1-p)^2$), then the DK model becomes site (resp. bond) DP. When $q=1$, the behavior of this model is different from those of the site DP and bond DP, as the latter ones belong to the so-called DP universality. Moreover, if $(p,q)=(1,0)$, then the DK model is equivalent to the well-known Wolfram Rule 90.

\section{Quantization of IPS \label{sec03}}
We consider a Markov chain corresponding to the IPS on a graph by regarding a configuration as a vertex of the graph. Let $G$ be a finite connected graph with $n$ vertices, $m$ nonloop edges and $n$ loops, where $G$ has one loop at each vertex of $G$. In this section, we consider $G$ as $G^{(i)}$ for $i=1,2$, so we assume that $G$ is connected. Furthermore, let $D_G $ be the symmetric digraph of $G$, where $D_G $ has one diloop at each vertex. Let $D(G)$ be the arc set of $D_G $. 
Then we have 
\begin{align*}
D(G)= \{ (u,u) \mid u \in V(G) \} \cup \{ (u,v), (v,u) \mid uv \in E(G), \ u \neq v \}, 
\end{align*}
where $V(G)$ is the set of vertices and $E(G)$ is the set of nonoriented edges $uv$ joining two vertices $u$ and $v$. For an arc $e=(u,v)$, set $o(e)=u $ and $t(e)=v$. The arc $e^{-1} =(v,u)$ is called the {\em inverse} of $e$. Note that, if $e$ is a diloop, then $e^{-1} =e$. Next, let $p: D(G) \longrightarrow [0,1]$ be a transition probability function such that for each $u \in V(G)$,  
\begin{align}
\sum_{o(e)=u} p(e)=1.
\label{wh01}
\end{align}
Then $(D_G , p)$ denotes a Markov chain defined by $p$ on $G$. The {\em transition matrix} ${\bf P} =(P_{uv} )_{u,v \in V(G)} $ is given by 
\[
P_{uv} =\left\{
\begin{array}{ll}
p(e) & \mbox{if $e=(u,v ) \in D(G)$, } \\
0 & \mbox{otherwise. }
\end{array}
\right.
\] 
Furthermore, the $(n+2m) \times (n+2m)$ matrix ${\bf U} =( U_{ef} )_{e,f \in D(G)} $ is defined as follows: 
\[
U_{ef} =\left\{
\begin{array}{ll}
2 \sqrt{p(e) p(f^{-1} )} & \mbox{if $t(f)=o(e)$ and $f \neq e^{-1} $, } \\
2 \sqrt{p(e) p(f^{-1} )}-1 & \mbox{if $f=e^{-1} $, } \\
0 & \mbox{otherwise. }
\end{array}
\right.
\] 
Note that ${\bf U} $ is unitary (see Proposition 1). The matrix ${\bf U} $ is called the {\em quantization} of $(D_G ,p)$ in this paper. 

Let ${\bf K} =( K_{ev} )_{e \in D(G); v \in V(G)} $ and ${\bf L} =( L_{ev} )_{e \in D(G); v \in V(G)} $ be the $(n+2m) \times n$ matrices given by 
\[
K_{ev} = \left\{
\begin{array}{ll}
\sqrt{p(e)} & \mbox{if $o(e)=v$, } \\
0 & \mbox{otherwise, } 
\end{array}
\right.
\ 
L_{ev} =\left\{
\begin{array}{ll}
\sqrt{p(e^{-1} )} & \mbox{if $t(e)=v$, } \\
0 & \mbox{otherwise. } 
\end{array}
\right.
\] 
Furthermore, we define an $(n+2m) \times (n+2m) $ matrix ${\bf J} =( J_{ef} )_{e,f \in D(G)} $ by
\[
J_{ef} =\left\{
\begin{array}{ll}
1 & \mbox{if $f=e^{-1}$ and $e$ is nonloop, } \\
1 & \mbox{if $f=e$ is a diloop, } \\
0 & \mbox{otherwise. } 
\end{array}
\right.
\] 
Then we get 
\begin{align}
{\bf U} =2 {\bf K} {\bf L}^T - {\bf J}. 
\label{wh02}
\end{align}

From now on, we show that ${\bf U}$ is unitary. 
\vspace*{12pt}
\par
\noindent
{\bf Proposition~1:}
Let $G$ be a connected graph with $n$ vertices, $m$ nonloop edges and $n$ loops, where $G$ has one loop at each vertex. Furthermore, let $(D_G ,p)$ be a Markov chain on $G$. Then the matrix ${\bf U} $ is unitary. 
\vspace*{12pt}
\par\noindent
{\bf Proof:} We begin with  
\begin{align}
{\bf L} = {\bf J} {\bf K}, \quad {\bf K} = {\bf J} {\bf L} . 
\label{hiro01}
\end{align}
Moreover, we get
\begin{align}
{\bf K}^T {\bf K} = {\bf I}_n , \quad {\bf J}^2 = {\bf I}_{n+2m} , \quad  {\bf J}^T = {\bf J} . 
\label{hiro02}
\end{align}
From Eqs. \eqref{hiro01} and \eqref{hiro02}, we see 
\begin{align}
{\bf L}^T {\bf L} = {\bf K}^T {\bf J}^T {\bf J} {\bf K} 
= {\bf K}^T {\bf J}^2 {\bf K} = {\bf K}^T {\bf K} = {\bf I}_{n} .
\label{hiro03}
\end{align}
Therefore, by using Eqs. \eqref{hiro01}, \eqref{hiro02} and \eqref{hiro03}, we obtain
\begin{align*}
{\bf U} {\bf U}^T &=  (2 {\bf K} {\bf L}^T - {\bf J} )(2 {\bf L} {\bf K}^T - {\bf J} ) 
\\
&=  4 {\bf K} {\bf L}^T {\bf L} {\bf K}^T -2 {\bf K} {\bf L}^T {\bf J} -2 {\bf J} {\bf L} {\bf K}^T + {\bf J}^2 
\\ 
&=  4 {\bf K} {\bf K}^T -2 {\bf K} {\bf K}^T -2 {\bf K} {\bf K}^T + {\bf I}_{n+2m} 
\\
& = {\bf I}_{n+2m}.
\end{align*}
Similarly, we get 
\begin{align*}
{\bf U}^T {\bf U} &= (2 {\bf L} {\bf K}^T - {\bf J} )(2 {\bf K} {\bf L}^T - {\bf J} ) 
\\
&= 4 {\bf L} {\bf K}^T {\bf K} {\bf L}^T -2 {\bf L} {\bf K}^T {\bf J} -2 {\bf J} {\bf K} {\bf L}^T + {\bf J}^2 
\\ 
&=  4 {\bf L} {\bf L}^T -4 {\bf L} {\bf L}^T + {\bf I}_{n+2m} 
\\
&=  {\bf I}_{n+2m}. 
\end{align*}
Hence, ${\bf U}$ is unitary. $\square$

\section{Zeta Function of Our Quantization \label{sec04}}
First we define a zeta function of $G$ with respect to ${\bf U}$ as follows: 
\[
\zeta (G, {\bf U} ,u)= \det ( {\bf I}_{n+2m} -u {\bf U} )^{-1} . 
\] 
In this section, we also consider $G$ as $G^{(i)}$ for $i=1,2$. Thus, $G$ is a connected graph. Then ${\bf U}$ corresponds to ${\bf U}^{(i)}$  for $i=1,2$. From now on, we present its determinant expression. The $n \times n$ matrix ${\bf S}=( S_{uv} )_{u,v \in V(G)}$ is determined by
\[
S_{uv} =\left\{
\begin{array}{ll}
\sqrt{p(e) p(e^{-1} )} & \mbox{if $e=(u,v) \in D(G)$, } \\
0 & \mbox{otherwise. }
\end{array}
\right.
\] 
Note that ${\bf S}$ is a symmetric matrix. Then the determinant expression for the zeta function $\zeta (G, {\bf U} ,u)$ is given in the following way. This is one of our main results.
\vspace*{12pt}
\par
\noindent
{\bf Theorem~2:}
Let $G$ be a connected graph with $n$ vertices, $m$ nonloop edges and $n$ loops, where $G$ has one loop at each vertex. Furthermore, let $(D_G ,p)$ be a Markov chain on $G$. Then we have  
\[
\zeta (G, {\bf U} ,u)^{-1} = (1+u)^n (1- u^2 )^{m-n} \det \left( (1+ u^2 ) {\bf I}_{n} -2u {\bf S} \right) . 
\]
\label{th1}
\vspace*{12pt}
\par
\noindent
{\bf Proof:} At first, we see 
\begin{align*}
\zeta \left( G, {\bf U}, u \right)^{-1} & =  \det \left( {\bf I}_{n+2m} -u {\bf U} \right) \\
&=  \det \left( {\bf I}_{n+2m} -u(2 {\bf K} {\bf L}^T - {\bf J} ) \right) \\
&=  \det \left( {\bf I}_{n+2m} +u {\bf J} -2u {\bf K} {\bf L}^T \right) \\
&=  \det \left( {\bf I}_{n+2m} -2u {\bf K} {\bf L}^T ({\bf I}_{n+2m} +u {\bf J})^{-1} \right) \det \left({\bf I}_{n+2m} +u {\bf J} \right). 
\end{align*}
The second equality comes from Eq. \eqref{wh02}. Here we should remark that if ${\bf A}$ and ${\bf B}$ are an $r \times s$ and $s \times r$ 
matrices, respectively, then we have 
\[
\det ( {\bf I}_{r} - {\bf A} {\bf B} ) = 
\det ( {\bf I}_s - {\bf B} {\bf A} ). 
\] 
Therefore, we have 
\begin{align}
\zeta (G, {\bf U} ,u)^{-1} = \det ( {\bf I}_{n} -2u {\bf L}^T ({\bf I}_{n+2m} +u {\bf J})^{-1} {\bf K} ) 
\det ({\bf I}_{n+2m} +u {\bf J}) . 
\label{hiro04}
\end{align}

Now, let $V(G)= \{ v_1 , \ldots, v_n \} $ and $D(G)= \{ e_1 , \ldots e_n , f_1 , f^{-1}_1 , \ldots , f_m , f^{-1}_m \} $, 
where $e_i =(v_i , v_i )$ is the diloop at each vertex $v_i \ (1 \leq i \leq n)$. 
Furthermore, put
\[ 
{\bf M} =
\left[ 
\begin{array}{cc}
0 & 1\\ 
1 & 0  
\end{array} 
\right] 
. 
\]
Then we have 
\[ 
{\bf J} = \left[ 
\begin{array}{cccc}
{\bf I}_n &  &  & {\bf 0} \\  
  & {\bf M} &  &  \\
  &   & \ddots &  \\
{\bf 0} &  &  & {\bf M} 
\end{array} 
\right] 
. 
\]
Thus, we obtain 
\begin{align}
\det ({\bf I}_{n+2m} +u {\bf J} )= \det 
\left[ 
\begin{array}{cccccc} 
(1+u) {\bf I}_n &  &  &  &  & {\bf 0} \\ 
  & 1 & u &  &  &  \\
  & u & 1 &  &  &  \\
  &   &   & \ddots &  &  \\ 
  &   &   &  & 1 & u \\
{\bf 0} &   &   &  & u & 1 
\end{array} 
\right] 
=(1+u)^n (1- u^2 )^m . 
\label{hiro05}
\end{align}
Moreover, we see 
\begin{align*}
&\left( {\bf I}_{n+2m} +u {\bf J} \right)^{-1} 
\\
& \qquad = 
\left[ 
\begin{array}{cccccc} 
(1+u) {\bf I}_n &  &  &  &  & {\bf 0} \\ 
  & 1 & u &  &  &  \\
  & u & 1 &  &  &  \\
  &   &   & \ddots &  &  \\ 
  &   &   &  & 1 & u \\
{\bf 0} &   &   &  & u & 1 
\end{array} 
\right]^{-1} 
\\ 
\\
& \qquad =  
\left[ 
\begin{array}{cccccc} 
1/(1+u) {\bf I}_n &  &  &  &  & {\bf 0} \\ 
  & 1/(1- u^2 ) & -u/(1-u^2 ) &  &  &  \\
  & -u/(1-u^2 ) & 1/(1- u^2 ) &  &  &  \\
  &   &   & \ddots &  &  \\ 
  &   &   &  & 1/(1- u^2 ) & -u/(1- u^2 ) \\
{\bf 0} &   &   &  &  -u/(1- u^2 ) & 1/(1- u^2 )  
\end{array} 
\right].  
\end{align*}

The $2m \times 2m$ matrix ${\bf J}_0 =( J^{(0)}_{ef} )_{e,f \in D(G); e,f \neq diloop} $ is given as follows: 
\[ 
J^{(0)}_{ef} =\left\{
\begin{array}{ll}
1 & \mbox{if $f=e^{-1}$, } \\
0 & \mbox{otherwise. } 
\end{array}
\right. 
\] 
Then we have 
\begin{align}
({\bf I}_{n+2m} +u {\bf J} )^{-1} = \frac{1}{1+u} {\bf I}_n \oplus \frac{1}{1- u^2 } ( {\bf I}_{2m} -u {\bf J}_0 ) ,  
\label{hiro06}
\end{align}
where ${\bf A} \oplus {\bf B} $ is the block diagonal sum of two square matrices ${\bf A} $ and ${\bf B} $. 
Therefore, combining Eq. \eqref{hiro04} with Eqs. \eqref{hiro05} and \eqref{hiro06} implies
\begin{align}
\zeta (G, {\bf U} ,u)^{-1} =(1+u)^n (1- u^2 )^m  \det \left[ {\bf I}_{n} - \frac{2u}{1- u^2 } {\bf L}^T \left\{ (1-u) {\bf I}_{n} \oplus ( {\bf I}_{2m} -u {\bf J}_0 ) \right\} {\bf K} \right]. 
\label{eq4}
\end{align}

Next, we put
\[ 
{\bf K} = 
\left[ 
\begin{array}{c}
{\bf K}_1 \\ 
{\bf K}_2 
\end{array} 
\right] 
, \qquad
{\bf L} = 
\left[ 
\begin{array}{c}
{\bf L}_1 \\ 
{\bf L}_2 
\end{array} 
\right]
, 
\]
where ${\bf K}_1$ and ${\bf L}_1$ are $n \times n$ matrices with respect to diloops $e_1 , \ldots , e_n $ and vertices $v_1 , \ldots , v_n $, and ${\bf K}_2$, and ${\bf L}_2$ are $2m \times n$ matrices with respect to nonloops $f_1 , f^{-1}_1 , \ldots , f_m ,f^{-1}_m $ and vertices $v_1 , \ldots , v_n $. Then we have 
\begin{align*}
{\bf L}^T ((1-u) {\bf I}_{n} \oplus {\bf I}_{2m} ) {\bf K}
& = 
\left[ 
\begin{array}{cc}
{\bf L}^T_1 & {\bf L}^T_2 
\end{array} 
\right] 
((1-u) {\bf I}_{n} \oplus {\bf I}_{2m} ) 
\left[ 
\begin{array}{c}
{\bf K}_1 \\ 
{\bf K}_2 
\end{array} 
\right] 
\\ 
&=  
\left[ 
\begin{array}{cc}
(1-u) {\bf L}^T_1 & {\bf L}^T_2 
\end{array} 
\right] 
\left[ 
\begin{array}{c}
{\bf K}_1 \\ 
{\bf K}_2 
\end{array} 
\right] 
\\ 
&= (1-u) {\bf L}^T_1 {\bf K}_1 + {\bf L}^T_2 {\bf K}_2. 
\end{align*}
Let ${\bf Q} = {\bf L}^T_1 {\bf K}_1 =( q_{uv} )_{u,v \in V(G)} $. 
Then we get 
\[
q_{uv} =\left\{
\begin{array}{ll}
p(e_u ) & \mbox{if $e_u = (u,v)$ and $u=v$, } \\
0 & \mbox{otherwise. } 
\end{array}
\right. 
\] 
Thus, we see that
\begin{align}
{\bf L}^T ((1-u) {\bf I}_{n} \oplus {\bf I}_{2m} ) {\bf K} = (1-u) {\bf L}^T_1 {\bf K}_1 + {\bf L}^T_2 {\bf K}_2 = (1-u) {\bf Q} + {\bf L}^T_2 {\bf K}_2,
\label{kouhaku01}
\end{align} 
since the second equality comes from ${\bf Q} = {\bf L}^T_1 {\bf K}_1$. Furthermore, we put 
\begin{align}
{\bf T} = {\bf L}^T_1 {\bf K}_1 + {\bf L}^T_2 {\bf K}_2 = {\bf Q} + {\bf L}^T_2 {\bf K}_2 . 
\label{kouhaku02}
\end{align}
Then we have 
\[
( {\bf T} )_{uv} =\left\{
\begin{array}{ll}
\sqrt{p(e) p(e^{-1} )} & \mbox{if $e=(v,u) \in D(G)$, } \\
0 & \mbox{otherwise. } 
\end{array}
\right. 
\] 
Note that $( {\bf T} )_{uu} =p(e_u )$ for $e_u = (u,u)$ \ $(u \in V(G))$. Thus, Eq. \eqref{kouhaku02} gives 
\begin{align}
{\bf L}^T_2 {\bf K}_2 = {\bf T} - {\bf Q}. 
\label{kouhaku03}
\end{align}
Therefore, combining Eq. \eqref{kouhaku01} with Eq. \eqref{kouhaku03} implies
\begin{align*}  
{\bf L}^T ((1-u) {\bf I}_{n} \oplus {\bf I}_{2m} ) {\bf K} 
= (1-u) {\bf Q} + \left( {\bf T} - {\bf Q} \right) = {\bf T} -u {\bf Q}.
\end{align*} 
That is, 
\begin{align}
{\bf L}^T ((1-u) {\bf I}_{n} \oplus {\bf I}_{2m} ) {\bf K}={\bf T} -u {\bf Q}.
\label{eq5} 
\end{align} 
Next, we obtain 
\begin{align*}
{\bf L}^T ( {\bf 0}_{n} \oplus (-u {\bf J}_0 )) {\bf K} 
& = 
\left[ 
\begin{array}{cc}
{\bf L}^T_1 & {\bf L}^T_2 
\end{array} 
\right] 
 ( {\bf 0}_{n} \oplus (-u {\bf J}_0 )) 
\left[ 
\begin{array}{c}
{\bf K}_1 \\ 
{\bf K}_2 
\end{array} 
\right] 
\\ 
&=  -u 
\left[ 
\begin{array}{cc}
{\bf 0} & {\bf L}^T_2 {\bf J}_0 
\end{array} 
\right] 
\left[ 
\begin{array}{c}
{\bf K}_1 \\ 
{\bf K}_2 
\end{array} 
\right] 
\\ 
&= -u {\bf L}^T_2 {\bf J}_0 {\bf K}_2. 
\end{align*}
Thus, we have 
\begin{align}
{\bf L}^T ( {\bf 0}_{n} \oplus (-u {\bf J}_0 )) {\bf K}=-u {\bf L}^T_2 {\bf J}_0 {\bf K}_2.  
\label{eq5b} 
\end{align} 
Therefore, combining Eq. \eqref{eq5} with Eq. \eqref{eq5b} implies 
\begin{align}
{\bf L}^T \left\{ (1-u) {\bf I}_{n} \oplus ( {\bf I}_{2m} -u {\bf J}_0 ) \right\} {\bf K} = {\bf T} -u \left( {\bf Q} + {\bf L}^T_2 {\bf J}_0 {\bf K}_2 \right).
\label{eq5c}
\end{align}

Here, ${\bf L}^T_2 {\bf J}_0 {\bf K}_2$ is a diagonal matrix and its $(u,u)$-entry for each vertex $u \in V(G)$ is 
\[
\sum_{o(e)=u, e \neq diloop} p(e) . 
\]
Thus, 
\begin{align}
{\bf Q} + {\bf L}^T_2 {\bf J}_0 {\bf K}_2
=  
\left[ 
\begin{array}{ccc}
\sum_{o(e)= v_1 } p(e) &  & 0 \\
  & \ddots &  \\
0 &  & \sum_{o(e)= v_n } p(e)  
\end{array} 
\right] 
={\bf I}_n. 
\label{eq6} 
\end{align}
The second equality comes from Eq. \eqref{wh01}. By using Eqs. \eqref{eq4}, \eqref{eq5c}, \eqref{eq6} and the fact that ${\bf S} = {\bf T}^T $, we obtain 
\begin{align*}
\zeta (G, {\bf U} ,u)^{-1} 
&=(1+u)^n (1- u^2 )^m  \det \left[ {\bf I}_{n} - \frac{2u}{1- u^2 } {\bf L}^T \left\{ (1-u) {\bf I}_{n} \oplus ( {\bf I}_{2m} -u {\bf J}_0 ) \right\} {\bf K} \right]
\\
&=(1+u)^n (1- u^2 )^m  \det \left[ {\bf I}_{n} - \frac{2u}{1- u^2 } 
\left\{ {\bf T} -u \left( {\bf Q} + {\bf L}^T_2 {\bf J}_0 {\bf K}_2 \right) \right\} \right]
\\
&=  (1+u)^n (1- u^2 )^m  \det \left[ {\bf I}_{n} - \frac{2u}{1- u^2 } 
\left( {\bf T} - u {\bf I}_n \right) \right] 
\\
&=  (1+u)^n (1- u^2 )^{m-n} \det \left[ (1+ u^2 ) {\bf I}_{n} -2u {\bf T} \right] 
\\ 
&=  (1+u)^n (1- u^2 )^{m-n} \det \left[(1+ u^2 ) {\bf I}_{n} -2u {\bf S} \right]. 
\end{align*}
$\square$

Note that, in general, the transition probability matrix ${\bf P} $ is not equal to the corresponding matrix ${\bf S} $. Substituting $u=1/ \lambda$ in Theorem 2, we obtain the characteristic polynomial for the matrix ${\bf U} $. 
\vspace*{12pt}
\par
\noindent
{\bf Corollary~3:}
Let $G$ be a connected graph with $n$ vertices, $m$ nonloop edges and $n$ loops, where $G$ has one loop at each vertex. Furthermore, let $(D_G ,p)$ be a Markov chain on $G$. Then we have 
\[
\det ( \lambda {\bf I}_{n+2m} - {\bf U} )=( \lambda +1)^n ( \lambda {}^2 -1)^{m-n} \det (( \lambda {}^2 +1) {\bf I}_{n} -2 \lambda {\bf S} ) . 
\]
\vspace*{12pt}
\par
From Corollary 3, we get eigenvalues for matrix ${\bf U} $. 
\vspace*{12pt}
\par
\noindent
{\bf Corollary~4:}
Let $G$ be a connected graph with $n$ vertices, $m$ nonloop edges and $n$ loops, where $G$ has one loop at each vertex. Furthermore, let $(D_G ,p)$  be a Markov chain on $G$. Then eigenvalues of ${\bf U}$ are given as follows: 

\vspace*{12pt}
\par\noindent
$\ $ (i) $2n$ eigenvalues: $\lambda = \mu \pm i \sqrt{1- \mu {}^2 } , \ \mu \in {\rm Spec} ( {\bf S} )$;  
\par\noindent
$\ $ (ii) $m$ eigenvalues: \ -1; 
\par\noindent
$\ $ (iii) $m-n$ eigenvalues: \ 1, 
\vspace*{12pt}
\par\noindent
where ${\rm Spec} (A)$ is the set of eigenvalues of a square matrix $A$.
\vspace*{12pt}
\par
\noindent
{\bf Proof:} By Corollary 3, we get 
\[
\det ( \lambda {\bf I}_{n+2m} - {\bf U} )=( \lambda +1)^n ( \lambda {}^2 -1)^{m-n} 
\prod_{\mu \in {\rm Spec} ( {\bf S} )} ( \lambda {}^2 +1-2 \mu \lambda ). 
\]
Solving $\lambda {}^2 -2 \mu \lambda +1=0$ gives $\lambda = \mu \pm i \sqrt{1- \mu {}^2 }$. 
$\square$

\section{DK Model \label{sec05}} 
First, we recall the local operator $Q^{(\ell)}_{DK} $ of the DK model: 
\[
Q^{(\ell)}_{DK} = 
\left[ 
\begin{array}{cccc}
1 & \cdot & 1-p & \cdot \\ 
\cdot & 1-p & \cdot & 1-q \\
\cdot & \cdot & p & \cdot \\
\cdot & p  & \cdot & q   
\end{array} 
\right] 
,  
\]
where $0 \leq p, \ q \leq 1$ and the rows and the columns correspond to $00$, $01$, $10$, $11$. We construct a new graph $G$ with four vertices  $00$, $01$, $10$, $11$ as follows: $uv \in E(G)$ if and only if the $(u,v)$-entry of $Q^{(\ell)}_{DK} $ is not zero. The order of vertices of $G$ is given as follows: $00, \ 10, \ 01, \ 11$. Then the local operator $Q^{(\ell)}_{DK}$ becomes
\[
Q^{(\ell)}_{DK} = 
\left[ 
\begin{array}{cccc}
1 & 1-p & \cdot & \cdot \\ 
\cdot & p & \cdot & \cdot \\
\cdot & \cdot & 1-p & 1-q \\
\cdot & \cdot & p  & q   
\end{array} 
\right] 
. 
\]

From now on, we consider $N=2$ case. Therefore, we see that ${\bf P} = Q^{(\ell)}_{DK}$. Then, let $G^{(1)}$ and $G^{(2)}$ be the complete subgraphs with vertex sets $\{00, 10\}$ and $\{01, 11\}$, respectively. It is clear that $G=G^{(1)} \cup G^{(2)}$ or $G$ is a disjoint union of $G^{(1)}$ and $G^{(2)}$. In this setting, we see  
\begin{align*}
{\bf P} 
&= Q^{(\ell)}_{DK} = {\bf P}^{(1)} \oplus {\bf P}^{(2)} = 
\left[ 
\begin{array}{cccc}
1 & 1-p & \cdot & \cdot \\ 
\cdot & p & \cdot & \cdot \\
\cdot & \cdot & 1-p & 1-q \\
\cdot & \cdot & p  & q   
\end{array} 
\right] 
, \qquad 
\\
{\bf P}^{(1)} 
&=
\left[ 
\begin{array}{cc}
1 & 1-p \\
\cdot & p 
\end{array} 
\right], 
\qquad 
{\bf P}^{(2)} 
=
\left[  
\begin{array}{cc}
1-p & 1-q \\
p & q 
\end{array} 
\right]
. 
\end{align*}
Moreover, ${\bf U}^{(i)}$ denotes the unitary matrix with respect to the quantization of $\left( D_{G^{(i)}} , p \right)$ for $ i=1,2$. Remark that ${\bf U}= {\bf U}^{(1)} \oplus {\bf U}^{(2)}$ is the unitary matrix with respect to the quantization of $\left( D_{G} , p \right)$ and ${\bf U}^{(i)}$ corresponds to ${\bf P}^{(i)}$ for $i=1,2$. Then we obtain 
\begin{align*}
{\bf U} 
&= {\bf U}^{(1)} \oplus {\bf U}^{(2)} =
\left[ 
\begin{array}{cc}
{\bf U}^{(1)} & O_2 \\ 
O_2 & {\bf U}^{(2)}    
\end{array} 
\right] 
, \qquad 
\\
{\bf U}^{(1)} 
&=
\left[ 
\begin{array}{cccc}
1 & \cdot & \cdot & \cdot \\
\cdot & \cdot & -1 & \cdot \\
\cdot & 1-2p & \cdot & 2 \sqrt{p(1-p)} \\
\cdot & 2 \sqrt{p(1-p)} & \cdot & 2p-1 
\end{array} 
\right], 
\\
{\bf U}^{(2)} 
&=
\left[ 
\begin{array}{cccc}
1-2p & \cdot & 2 \sqrt{p(1-p)} & \cdot \\
2 \sqrt{p(1-p)} & \cdot & 2p-1 & \cdot \\
\cdot & 1-2q & \cdot & 2 \sqrt{q(1-q)} \\
\cdot & 2 \sqrt{q(1-q)} & \cdot & 2q-1 
\end{array} 
\right] 
, 
\end{align*}
where $O_n$ is the $n \times n$ zero matrix. Furthermore, the symmetric matrix ${\bf S}$ is obtained by ${\bf P} = Q^{(\ell)}_{DK}$ as follows:.
\[
{\bf S} =
\left[ 
\begin{array}{cccc}
1 & \cdot & \cdot & \cdot \\
\cdot & p & \cdot & \cdot \\
\cdot & \cdot & 1-p & \sqrt{p(1-q)} \\
\cdot & \cdot & \sqrt{p(1-q)} & q  
\end{array} 
\right] 
. 
\]
The characteristic polynomials for the matrices ${\bf P}^{(1)} $, ${\bf P}^{(2)} $, ${\bf U}^{(1)} $, and ${\bf U}^{(2)} $ can be derived from the direct computation in the following way.
\begin{align*}
\det ( x {\bf I}_2 - {\bf P}^{(1)} ) 
&=(x-1) (x-p),  
\\
\det ( x {\bf I}_2 - {\bf P}^{(2)} ) 
&=(x-1) (x-q+p),  
\\
\det ( x {\bf I}_4 - {\bf U}^{(1)} ) 
&=(x-1)(x+1)( x^2 -2px+1),   
\\
\det ( x {\bf I}_4 - {\bf U}^{(2)} )
&=(x-1)(x+1)( x^2 +2(p-q)x+1).   
\end{align*}
Thus, we get 
\begin{align*}
\det ( x {\bf I}_4 - {\bf P} ) 
&=\det ( x {\bf I}_2 - {\bf P}^{(1)} ) \det ( x {\bf I}_2 - {\bf P}^{(2)} ) = (x-1)^2 (x-p)(x-q+p),
\\
\det ( x {\bf I}_8 - {\bf U} ) 
&=\det ( x {\bf I}_4 - {\bf U}^{(1)} ) \det ( x {\bf I}_4 - {\bf U}^{(2)} ) 
\\
&= (x-1)^2 (x+1)^2 ( x^2 -2px+1)( x^2 +2(p-q)x+1).
\end{align*}
In particular, we have
\begin{align}
\det ( x {\bf I}_8 - {\bf U} ) = (x-1)^2 (x+1)^2 ( x^2 -2px+1)( x^2 +2(p-q)x+1).\label{hiro001}
\end{align}
Furthermore, 
\begin{align}
\det ( x {\bf I}_4 - {\bf S} )=(x-1)^2 (x-p)(x-q+p) .   
\label{wata01}
\end{align}
On the other hand, combining Corollary 3 with Eq. \eqref{wata01} implies 
\begin{align*}
&\det \left( x {\bf I}_8 - {\bf U} \right) 
\\
&= (x+1)^4 (x^2 -1)^{2-4} \det \left(( x^2 +1) {\bf I}_4 -2x {\bf S} \right) 
\\
&= (x+1)^4 (x^2 -1)^{-2} (2x)^4 \det \left( \frac{x^2 +1}{2x} {\bf I}_4 - {\bf S} \right) 
\\
&= (x+1)^4 (x -1)^{-2} (x+1)^{-2} \cdot 16 x^4 \left( \frac{x^2 +1}{2x} -1 \right)^2 \left( \frac{x^2 +1}{2x} -p \right) \left( \frac{x^2 +1}{2x} -q+p \right) 
\\
&= (x+1)^2 (x -1)^2 (x^2 -2px+1) (x^2 +2(p-q)x+1). 
\end{align*}
Then we have the same expression for $\det \left( x {\bf I}_8 - {\bf U} \right)$ given in Eq. \eqref{hiro001}. We should remark that ${\bf P} \neq {\bf S}$, however $\det (x {\bf I}_4 - {\bf P} )= \det (x {\bf I}_4 - {\bf S} )$. One of the future interesting problems is to consider the relation between $\det (x {\bf I}_{2^N} - {\bf P} )$ and $\det (x {\bf I}_{2^N} - {\bf S} )$ for $N \ge 3$, since we have $\det (x {\bf I}_{2^3} - {\bf P} ) \not= \det (x {\bf I}_{2^3} - {\bf S} )$ for some $(p,q)$ when $N = 3$. To be more precise, if $(p,q)=(1/3,1/2)$, then we see that $\det (x {\bf I}_{2^2} - {\bf P}^{(1)} ) = \det (x {\bf I}_{2^2} - {\bf S}^{(1)} )$ and $\det (x {\bf I}_{2^2} - {\bf P}^{(2)} ) \not= \det (x {\bf I}_{2^2} - {\bf S}^{(2)} )$, where ${\bf S}^{(i)}$ is the symmetrized matrix of ${\bf U}^{(i)}$ for $i=1,2$ with ${\bf S} = {\bf S}^{(1)} \oplus {\bf S}^{(2)}$.

\section{Absolute Zeta Function \label{sec06}} 
In this section, we briefly review the framework on the absolute zeta functions, which can be considered as zeta function over $\mathbb{F}_1$, and absolute automorphic forms (see \cite{CC, KF1, Kurokawa3, Kurokawa, KO, KT3, KT4, Soule} and references therein, for example). 

Let $f(x)$ be a function $f : \mathbb{R} \to \mathbb{C} \cup \{ \infty \}$. We say that $f$ is an {\em absolute automorphic form} of weight $D$ if $f$ satisfies
\begin{align*}
f \left( \frac{1}{x} \right) = C x^{-D} f(x)
\end{align*}
with $C \in \{ -1, 1 \}$ and $D \in \mathbb{Z}$. The {\em absolute Hurwitz zeta function} $Z_f (w,s)$ is defined by
\begin{align*}
Z_f (w,s) = \frac{1}{\Gamma (w)} \int_{1}^{\infty} f(x) \ x^{-s-1} \left( \log x \right)^{w-1} dx,
\end{align*}
where $\Gamma (x)$ is the gamma function (see \cite{Andrews1999}, for instance). Then taking $x=e^t$, we see that $Z_f (w,s)$ can be rewritten as the Mellin transform: 
\begin{align*}
Z_f (w,s) = \frac{1}{\Gamma (w)} \int_{0}^{\infty} f(e^t) \ e^{-st} \ t^{w-1} dt.
\end{align*}
Moreover, the {\em absolute zeta function} $\zeta_f (s)$ is defined by 
\begin{align*}
\zeta_f (s) = \exp \left( \frac{\partial}{\partial w} Z_f (w,s) \Big|_{w=0} \right).
\end{align*}
Here we introduce the {\em multiple Hurwitz zeta function of order $r$}, $\zeta_r (s, x, (\omega_1, \ldots, \omega_r))$, the {\em multiple gamma function of order $r$}, $\Gamma_r (x, (\omega_1, \ldots, \omega_r))$, and the {\em multiple sine function of order $r$}, $S_r (x, (\omega_1, \ldots, \omega_r))$, respectively (see \cite{CC, Kurokawa3, Kurokawa, KT3}, for example): 
\begin{align*}
\zeta_r (s, x, (\omega_1, \ldots, \omega_r))
&= \sum_{n_1=0}^{\infty} \cdots \sum_{n_r=0}^{\infty} \left( n_1 \omega_1 + \cdots + n_r \omega_r + x \right)^{-s}, 
\\
\Gamma_r (x, (\omega_1, \ldots, \omega_r)) 
&= \exp \left( \frac{\partial}{\partial s} \zeta_r (s, x, (\omega_1, \ldots, \omega_r)) \Big|_{s=0} \right),
\\
S_r (x, (\omega_1, \ldots, \omega_r))
&= \Gamma_r (x, (\omega_1, \ldots, \omega_r))^{-1} \ \Gamma_r (\omega_1+ \cdots + \omega_r - x, (\omega_1, \ldots, \omega_r))^{(-1)^r}.
\end{align*}
\par
Now we present the following key result derived from Theorem 4.2 and its proof in Korokawa \cite{Kurokawa} (see also Theorem 1 in Kurokawa and Tanaka \cite{KT3}):

\vspace*{12pt}
\noindent
{\bf Theorem~5:}
For $\ell \in \mathbb{Z}, \ m(i) \in \mathbb{Z}_{>} \ (i=1, \ldots, a), \ n(j) \in \mathbb{Z}_{>} \ (j=1, \ldots, b)$, put 
\begin{align*}
f(x) = x^{\ell/2} \ \frac{\left( x^{m(1)} - 1 \right) \cdots \left( x^{m(a)} - 1 \right)}{\left( x^{n(1)} - 1 \right) \cdots \left( x^{n(b)} - 1 \right)}.
\end{align*}
Then we have 
\begin{align}
Z_f (w, s) 
&= \sum_{I \subset \{1, \ldots, a \}} (-1)^{|I|} \ \zeta_b \left( w, s - {\rm deg} (f) + m \left( I \right), \bec{n} \right),
\nonumber
\\
\zeta_f (s) 
&= \prod_{I \subset \{1, \ldots, a \}} \Gamma_b \left( s - {\rm deg} (f) + m \left( I \right), \bec{n} \right)^{ (-1)^{|I|}},
\nonumber
\\
\zeta_f \left( D-s \right)^C 
&= \varepsilon_f (s) \ \zeta_f (s),
\label{mkusatsufe}
\end{align}
where
\begin{align*}
|I| 
&= \sum_{i \in I} 1, \quad {\rm deg} (f) = \frac{\ell}{2} + \sum_{i=1}^a m(i)- \sum_{j=1}^b n(j), \quad m \left( I \right) = \sum_{i \in I} m(i),
\\
\bec{n} 
&= \left( n(1), \ldots, n(b) \right), \quad D = \ell + \sum_{i=1}^a m(i)- \sum_{j=1}^b n(j), \quad C=(-1)^{a-b}, 
\\
\varepsilon_f (s) 
&= \prod_{I \subset \{1, \ldots, a \}} S_b \left( s - {\rm deg} (f) + m \left( I \right), \bec{n} \right)^{ (-1)^{|I|}}.
\end{align*}

\vspace*{12pt}
\noindent
We should note that Eq. \eqref{mkusatsufe} is called the {\em functional equation}.

Finally, we consider $N=2$ case of the DK model. Note that for our IPS including the DK model, the zeta function $\zeta (G, {\bf U}, u)$ becomes an absolute automorphic form for any $N \ge 2$, since ${\bf U}$ is an orthogonal matrix. This argument can be seen in our previous papers \cite{Konno2023, AkahoriEtAL2023} where we computed the absolute zeta function for zeta functions based on quantum walk \cite{Konno2023} and QCA \cite{AkahoriEtAL2023}.

If $(p,q)=(1/2,0)$ or $(p,q)=(0,1/2)$, then we can calculate the absolute zeta function of the zeta function $\zeta (G, {\bf U}, u)$ by using Theorem 5. From now on, we put $\zeta_{{\bf U}} (u) = \zeta (G, {\bf U}, u)$ for short.
\par
\
\par
(i) $(p,q)=(1/2,0)$ case. By Eq. \eqref{hiro001}, we have
\begin{align*}
\det ( x {\bf I}_8 - {\bf U} ) 
= (x-1)^2 (x+1)^2 ( x^2 -x+1)( x^2 +x+1). 
\end{align*}
Substituting $x=1/u$ in the above equation, we get
\begin{align*}
\zeta_{{\bf U}} (u) = \det ( {\bf I}_8 - u {\bf U} )^{-1} = \frac{1}{(u^2-1)(u^6-1)}. 
\end{align*}
Noting that $\ell =0, \ a=0, \ b=2, \ n(1)=2, \ n(2)=6, \ {\rm deg} (\zeta_{{\bf U}} ) = D = -8$ and $C=1$, it follows from Theorem 5 that 
\begin{align*}
Z_{\zeta_{{\bf U}}} (w, s) 
&= \zeta_{2} \left(w, s + 8, (2,6) \right),
\\
\zeta_{\zeta_{{\bf U}}}(s) 
&= \Gamma_{2} \left( s + 8, (2,6) \right),
\\
\zeta_{\zeta_{{\bf U}}}(-8-s)
&= S_{2} \left( s + 8, (2,6) \right) \ \zeta_{\zeta_{{\bf U}}}(s).
\end{align*}
Thus, by the above functional equation, we have a critical value $s=-4$, since $-8-s=s$. Then $\zeta_{\zeta_{{\bf U}}}(s)$ at $s=-4$ is given by 
\begin{align*}
\zeta_{\zeta_{{\bf U}}}(-4) = \Gamma_{2} \left( 4, (2,6) \right).
\end{align*}
\par
\
\par
(ii) $(p,q)=(0,1/2)$ case. From Eq. \eqref{hiro001}, we see
\begin{align*}
\det ( x {\bf I}_8 - {\bf U} ) 
= (x-1)^2 (x+1)^2 ( x^2 +1)( x^2 -x+1). 
\end{align*}
Substituting $x=1/u$ in the above equation implies
\begin{align*}
\zeta_{{\bf U}} (u) = \det ( {\bf I}_8 - u {\bf U} )^{-1} = \frac{u^3-1}{(u-1)(u^4-1)(u^6-1)}. 
\end{align*}
Noting that $\ell =0, \ a=1, \ b=3, \ m(1)=3, \ n(1)=1, \ n(2)=4, \ n(3)=6, \ {\rm deg} (\zeta_{{\bf U}} ) = D = -8$ and $C=1$, by Theorem 5 we get
\begin{align*}
Z_{\zeta_{{\bf U}}} (w, s) 
&= \sum_{I \subset \{1\}} \zeta_{3} \left(s+ 8 + m(I), (1,4,6) \right),
\\
\zeta_{\zeta_{{\bf U}}}(s) 
&= \prod_{I \subset \{1\}} \Gamma_3 (s+ 8 + m(I), (1,4,6))^{(-1)^{|I|}},
\\
\zeta_{\zeta_{{\bf U}}}(-8-s)
&=  \left\{ \prod_{I \subset \{1\}} S_{3} \left( s + 8 + m(I), (1,4,6) \right)^{ (-1)^{|I|}} \right\} \ \zeta_{\zeta_{{\bf U}}}(s).
\end{align*}
Therefore, the above functional equation gives a critical value $s=-4$, since $-8-s=s$. Then we have  $\zeta_{\zeta_{{\bf U}}}(s)$ at $s=-4$ as follows:
\begin{align*}
\zeta_{\zeta_{{\bf U}}}(-4) = \frac{\Gamma_{3} \left( 4, (1,4,6) \right)}{\Gamma_{3} \left( 7, (1,4,6) \right)}.
\end{align*}

\section{Summary \label{sec07}}
In this paper, we considered a new quantization ${\bf U}$ of the Markov chain determined by ${\bf P}$ on a graph $G$. We introduced a zeta function $\zeta (G, {\bf U}, u)$ of $G$ for ${\bf U}$ and gave its determinant expression (Theorem 2). After that, we applied this quantization to IPSs which are PCA with nearest-neighbor interaction including the DK model. A special case of the DK model is directed percolation. Moreover, we calculated the absolute zeta function $\zeta_{\zeta_{{\bf U}}}(s)$ of a quantized model of the DK model for $N=2$ case. One of the interesting future problems might be to determine the critical line of the DK model for $N = \infty$ case by using the functional equation of the absolute zeta function for our zeta function of the quantized model of the DK model. 



\begin{thebibliography}{99}


\bibitem{DomanyKinzel1984} 
E. Domany and W. Kinzel (1984), 
{\it Equivalence of cellular automata to Ising models and directed percolation}, Phys. Rev. Lett., Vol.53, pp. 311-314.

\bibitem{Durett1988} 
R. Durrett (1988), 
{\it Lecture notes on particle systems and percolation}, 
Wadsworth, Inc.


\bibitem{Lubeck2006} 
S. L\"ubeck (2006), 
{\it Crossover scaling in the Domany-Kinzel cellular automaton}, 
J. Stat. Mech., Vol. 2006, P09009.


\bibitem{HHLP2008} 
M. Henkel, H. Hinrichsen and S. L\"ubeck (2008),
{\it Non-equilibrium phase transition}, 
Springer (Dordrecht).


\bibitem{Konno2008}
N. Konno (2008), 
{\it Quantum walks}, In: Quantum Potential Theory, Franz, U., Schurmann, M., Eds., Lecture Notes in Mathematics: Vol. 1954, pp. 309-452, Springer (Heidelberg).

\bibitem{Venegas} 
S.E. Venegas-Andraca (2012),  
{\it Quantum walks: a comprehensive review}, 
Quantum Inf. Process., Vol. 11, pp. 1015-1106.


\bibitem{Portugal} 
R. Portugal (2018), 
{\it Quantum walks and search algorithms, 2nd edition},  
Springer (New York).


\bibitem{KonnoSato} 
N. Konno and I. Sato (2012), 
{\it On the relation between quantum walks and zeta functions},  
Quantum Inf. Process., Vol. 11, pp. 341-349.


\bibitem{Konno2023}
N. Konno (2023),  
{\it On the relation between quantum walks and absolute zeta functions}, 
Quantum Stud.: Math. Found. (in press), arXiv:2306.14625.


\bibitem{AkahoriEtAL2023}
J. Akahori, N. Konno and I. Sato (2023), 
{\it Absolute zeta functions for zeta functions of quantum cellular automata}, 
Quantum Inf. Comput., Vol. 23, pp. 1261-1274. 


\bibitem{KomatsuEtAl2022} 
T. Komatsu, N. Konno and I. Sato (2022), 
{\it IPS/Zeta correspondence}, 
Quantum Inf. Comput., Vol. 22, pp. 251-269.


\bibitem{KiumiEtAl2022}
C. Kiumi, N. Konno and Y. Oshima (2022),
{\it IPS/Zeta correspondence for the Domany-Kinzel model}, 
arXiv:2206.03188. 


\bibitem{KatoriEtAl2004} 
M. Katori, N. Konno, A. Sudbury and H. Tanemura (2004), 
{\it Dualities for the Domany-Kinzel model}, 
J. Theoret. Probab., Vol. 17, pp. 131-144.


\bibitem{Konno2008b} 
N. Konno (2008), 
{\it Quantum walks and quantum cellular automata}, In: Cellular automata, H. Umeo, S. Morishita, K. Nishinari, T. Komatsuzaki and S. Bandini, Eds., Lecture Notes in Computer Science: Vol.5191, pp. 12-21, Springer (Heidelberg).


\bibitem{CC} 
A. Connes and C. Consani (2010),  
{\it Schemes over $\mathbb{F}_1$ and zeta functions},
Compositio Math., Vol. 146, pp. 1383-1415.


\bibitem{KF1}
N. Kurokawa (2005), 
{\it Zeta functions over $\mathbb{F}_1$}, 
Proc. Japan Acad. Ser. A \ Math. Sci., Vol. 81, pp. 180-184.


\bibitem{Kurokawa3}  
N. Kurokawa (2013), 
{\it Modern theory of trigonometric functions}, 
Iwanami Publication (Tokyo) in Japanese.


\bibitem{Kurokawa}  
N. Kurokawa (2016), 
{\it Theory of absolute zeta functions}, 
Iwanami Publication (Tokyo) in Japanese.


\bibitem{KO}
N. Kurokawa and H. Ochiai (2013),
{\it Dualities for absolute zeta functions and multiple gamma functions}, 
Proc. Japan Acad. Ser. A \ Math. Sci., Vol. 89, pp. 75-79. 


\bibitem{KT3}
N. Kurokawa and H. Tanaka (2017), 
{\it Absolute zeta functions and the automorphy}, 
Kodai Math. J., Vol. 40, pp. 584-614.


\bibitem{KT4}
N. Kurokawa and H. Tanaka (2018), 
{\it Absolute zeta functions and absolute automorphic forms}, 
J. Geom. Phys., Vol. 126, pp. 168-180.


\bibitem{Soule}
C. Soul\'e (2004), 
{\it Les vari\'et\'es sur le corps \`a un \'el\'ement}, 
Mosc. Math. J., Vol. 4, pp. 217--244.


\bibitem{Andrews1999} 
G. E. Andrews, R. Askey and  R. Roy (1999), 
{\it Special functions}, 
Cambridge University Press (New York).












\end{thebibliography}
\end{document}